\documentstyle[aps,pra,epsfig,twocolumn]{revtex} 
\def\bma{\begin{mathletters}}
\def\ema{\end{mathletters}}

\begin{document}

\title{Matrix models of 4-dimensional quantum Hall fluids}

\author{Yi-Xin Chen } 
\address{Zhejiang Institute of Modern Physics, Zhejiang University, 
             Hangzhou 310027, P. R.China}

\date{\today}
\maketitle

\begin{abstract}

We propose two non-abelian Chern-Simons matrix models as the effective descriptions of 4-dimensional
quantum Hall fluids. One of them describes a new type of 4-dimensional
quantum Hall fluid on the space of quaternions, the other provides the description of non-commutative field theory
for Zhang and Hu's 4-dimensional quantum Hall fluid of $S^4$. The complete sets of physical quantum states of these matrix models are determined, and the properties of quantum Hall fluids related with them are discussed.


\vspace{.3cm}


\end{abstract}

\setcounter{equation}{0}

\indent

There has recently appeared an interesting connection between quantum 
Hall effect (QHE)\cite{Laughlin} and non-commutative field theory. In particular, Susskind 
\cite{Susskind} proposed that non-commutative Chern-Simons theory on the 
plane may provide a description of the quantum Hall fluid (QHF). Susskind's non-commutative 
Chern-Simons theory on the plane describes a spatially infinite quantum 
Hall system. The fields of this theory are infinite matrices that act on 
an infinite Hilbert space, appropriate to account for an infinite number of 
electrons. Polychronakos \cite{Polychronakos} proposed a 
matrix regularized version of Susskind's non-commutative Chern-Simons 
theory in effort to describe finite systems with a finite number of 
electrons in the limited spatial extent. This matrix model was shown to 
reproduce the basic properties of the QHF\cite{Hellerman}. 

Recently, Zhang and Hu \cite{Zhang,Hu} have found a
4-dimensional generalization of the quantum Hall system that is composed of 
many particles moving in four dimensional space under a $SU(2)$ gauge 
field. The authors in \cite{Fabinger,Chen,Karabali1,Kimura}have developed 
the idea of Zhang and Hu in other directions.

The goal of this paper is to establish the second-quantized field theory description of 4-dimensional QHFs, 
i.e, to do the non-commutative field theories for 4-dimensional QHFs. 

The matrix models proposed in the present paper originate from the investigation of matrix formulation of M-theory \cite{Banks}.
In the matrix model under the background of the longitudinal M5-branes \cite{Berkooz}, the open membrane can be ended on the M5-brane similar to the fundamental string ending on the D-branes in the string picture. If one consider the rigid properties of
open membrane \cite{Berkooz1}, the gauge transformation of area preserving can lead to a non-commutative Chern-Simons 
theory \cite{Chen1}. For the bound configurations of the open membrane and the spherical 5-brane in the fixed background,
the non-commutative Chern-Simons field is valued in the Lie algebra $SO(4)$. The matrix models presented here are the matrix
regularized models of such non-commutative non-abelian Chern-Simons theories. Since the group $SO(4)$ is isomorphic to $SU(2)\times
SU(2)$, the $SU(2)$ finite matrix Chern-Simons model is the fundamental building block of these non-commutative Chern-Simons
theories. Hence, my starting point is the $SU(2)$ finite matrix Chern-Simons model. 

The action of this matrix model is
\begin{eqnarray}
S&=&\frac{H}{4}\int dt Tr\{ {\bar Z}^{\alpha} i(\partial_t Z^{\alpha} +[A_{0}^{\alpha\beta},Z^{\beta}])- 2i\theta A_{0} \nonumber\\
&-&\omega {\bar Z}^{\alpha}Z^{\alpha} \}
+\frac{1}{2}\int dt\Psi^{\dagger\alpha}i(\partial_t \Psi^{\alpha} +A_{0}^{\alpha\beta}\Psi^{\beta})+ h.c.
\end{eqnarray}
Here, we consider the system composed of $N$ particles. The complex coordinates $Z^{\alpha}$, which can be expressed as 
$Z^{1}=Q^1+iQ^2$ and $Z^{2}=Q^3+iQ^4$ in terms of four real coordinates, are represented by finite $N\times N$ matrices. The gauge fields $A_{0}^{\alpha\beta}$ obey the relations $A_{0}^{11}=-A_{0}^{22\dagger}$ and $A_{0}^{12}=-A_{0}^{21\dagger}$, and each element of them is a $N\times N$ matrix. The constant $H$ plays the role similar to the constant magnetic field in 2-dimensional QHE. $\theta$ is the positive parameter characterizing the non-commutativity of the coordinates of particles. The action (1) is invariant under the gauge transformations $Z^{\alpha}\rightarrow UZ^{\alpha}U^{-1}$, $\Psi^{\alpha}\rightarrow U\Psi^{\alpha}$ and $A_{0}^{\alpha\beta}\rightarrow UA_{0}^{\alpha\beta}U^{-1}-\delta^{\alpha\beta}\partial_t UU^{-1}$ if the level of this Chern-Simons theory is an integer. However, the $U(N)$ gauge transformations are combined with the $SU(2)$ gauge transformations (see below). By choosing the gauge $A_{0}^{\alpha\beta}=0$ and imposing the equation of motion of 
$ A_{0}^{\alpha\beta}$ as the constraints, we obtain
\bma
\begin{eqnarray}
\frac{H}{2}[Z^{1},{\bar Z}^{1}]&+&\frac{H}{2}[Z^{2},{\bar Z}^{2}]+\Psi^{1}\Psi^{1\dagger}+\Psi^{2}\Psi^{2\dagger}=2H\theta, \\
\frac{H}{2}[Z^{2},{\bar Z}^{1}]&+&\Psi^{2}\Psi^{1\dagger}=0.
\end{eqnarray}
\ema

We will proceed by first quantizing our theory before solving the constraints, and
then applying the constraints as operator conditions on the physical states. After performing the quantization 
of the system, one can find that the matrix elements of $Z^{\alpha}$ and the components 
of $\Psi^{\alpha}$ satisfy the following commutation relations
\begin{equation}
[Z^{\alpha}_{ij},Z^{\beta\dagger}_{kl}]=\delta^{\alpha\beta} \delta_{il} \delta_{jk},
[\Psi^{\alpha}_{i},\Psi^{\beta\dagger}_{j}]=\delta^{\alpha\beta} \delta_{ij},
\end{equation}
where we had rescaled $\sqrt{H/2}Z^{\alpha}$ and $\sqrt{H/2}{\bar Z}^{\alpha}$ into $Z^{\alpha}$ and ${\bar Z}^{\alpha}$ respectively. Then, one can easily write the hamiltonian of the matrix model at hand as
\begin{equation}
{\cal H}=\omega (N^{2}+Tr(Z^{\alpha\dagger}Z^{\alpha})).
\end{equation}
Following \cite{Polychronakos}, we can easily fix the operator expression of the Gauss constraints after the quantization. The operators of the traceless part of the constraints (2) lead to the conditions of physical states as
\begin{equation}
G^{a,3} | phys \rangle  =0,
G^{a,+} | phys \rangle  =0,
\end{equation}
where $G^{a,3}$ and $G^{a,+}$ are given by  
\begin{eqnarray}
G^{a,3}&=&z^{1\dagger}_{b}(-if^{abc})z^{1}_{c}
+\Psi^{1\dagger}_{i}T_{ij}^{a}\Psi^{1}_{j}\nonumber\\
&+&z^{2\dagger}_{b}(-if^{abc})z^{2}_{c}+\Psi^{2\dagger}_{i}T_{ij}^{a}\Psi^{2}_{j},
\end{eqnarray}
\begin{equation}
G^{a,+}=z^{1\dagger}_{b}(-if^{abc})z^{2}_{c}+\Psi^{1\dagger}_{i}T_{ij}^{a}\Psi^{2}_{j}.
\end{equation}
In (6) and (7), $z_{a}^{\alpha}$ represent the components of the expanding of $Z^{\alpha}$ in the $SU(N)$ basis of matrices. It can be easily proved that the generators of the constraints satisfy the commutation relations $[G^{a,3},G^{b,3}]=if^{abc}G^{c,3}$ and $[G^{a,3},G^{b,+}]=if^{abc}G^{c,+}$.
It should be pointed that if we choose the Gauss constraints (5), $G^{a,-}=(G^{a,+})^{\dagger}$ acting on the physical states does not produce the vanishing result. So the physical states are not the $SU(2)$ invariants, but they are the $SU(N)$ invariants. Taking the trace of (2a), we obtain
\begin{equation}
\left[ \Psi^{1\dagger}_{i}\Psi^{1}_{i} +\Psi^{2\dagger}_{i}\Psi^{2}_{i} -2NH\theta \right] |phys\rangle  =0 .
\end{equation}
The trace part of (2b) $\Psi^{1\dagger}_{i} \Psi^{2}_{i} =0$ gives the orthogonality of 
$\Psi^{1}$ and $\Psi^{2}$ as the $N$-vectors under the transformation of $SU(N)$. After quantization, its operator condition
implies that it is impossible that $\Psi^{1\dagger}$ together with $\Psi^{2\dagger}$ appear in the physical states in a singlet of $SU(N)$.

By considering the $SU(N)$ invariance of the physical states and the structure of (5), one can easily show that $2H\theta$ is an integer, 
i.e., $2H\theta=k$. Furthermore, the orthogonality 
of $\Psi^{1}$ and $\Psi^{2}$ make the equation (8) become 
$\left[ \Psi^{1\dagger}_{i}\Psi^{1}_{i}-Nk \right] |phys\rangle  =0$. Define the Fock space
vaccum $| 0  \rangle $ by $Z^{\alpha}_{ij}|0 \rangle =\Psi^{\alpha}_{i} |0 \rangle=0 $. Thus, any physical state describing an energy eigenstate of the hamiltonian (4) should be the linear combinations of the following Fock states
\begin{equation}
\prod_{m=1}^{M}(Z^{1\dagger})^{i_{m}}_{j_{m}} \prod_{m^{\prime}=1}^{M^{\prime}}(Z^{2\dagger})^{i_{m^{\prime}}^{\prime}}_{j_{m^{\prime}}^{\prime}}
\prod_{n=1}^{Nk}(\Psi^{1\dagger})_{l_{n}} 
 |0 \rangle , 
\end{equation}
where, the fundamental indices of $SU(N)$ are written as upper indices, and the anti-fundamental indices as lower indices.

Now, our task is to make the states of (9) into the singlets of $SU(N)$ by contracting all $SU(N)$ indices of it. We can do this in two manners.
One is to contract the indices of $\Psi^{1\dagger}$ and $Z^{\alpha\dagger}$ in the $SU(2)$ invariant way. The key point is to employ the
fundamental fact that $\prod_{i<j} (u_{i}v_{j} -u_{j}v_{i})=\epsilon^{i_{1}\cdots i_{N}}\prod_{n=1}^{N}(u^{N-n}v^{n-1})_{i_{n}}$, where $u$ and $v$ are two components of spinor representation of $SU(2)$. In this manner, we obtain the
fundamental block  
$\epsilon^{i_{1}\cdots i_{N}}\prod_{n=1}^{N}(\Psi^{1\dagger}Z^{1\dagger N-n}Z^{2\dagger n-1})_{i_{n}}$ in a singlet of $SU(N)$. Another manner is based on that $\Psi^{1\dagger},Z^{1\dagger}$ and $\Psi^{2\dagger},Z^{2\dagger}$ are the components of 'spin up' and those of 'spin down' respectively, since both $\Psi^{\alpha\dagger}$ and $Z^{\alpha\dagger}$
belong to the spinor representations of $SU(2)$. The $SU(N)$ invariant block are then given by 
$\epsilon^{i_{1}\cdots i_{N}}\prod_{m=1}^{N}(\Psi^{1\dagger}Z^{1\dagger n_{m}+1}Z^{2\dagger n_{m}})_{i_{m}}$. 
One can easily find that these fundamental blocks really satisfy the Gauss constraint conditions of the physical states.
By employing the blocks and the technique in \cite{Hellerman}, we can write the Fock basis of the physical states of the matrix model as
\begin{eqnarray}
|\{ c_{i} \} ,&\{ &c_{i}^{\prime} \}, k \rangle =\prod_{j=1}^{N}(Tr Z^{1\dagger j})^{c_{j}}(Tr Z^{2\dagger j})^{c_{j}^{\prime}} \nonumber\\
&(&\epsilon^{i_{1}\cdots i_{N}}\prod_{n=1}^{N}(\Psi^{1\dagger}Z^{1\dagger N-n}Z^{2\dagger n-1})_{i_{n}}
)^{k}|0 \rangle  ,
\end{eqnarray}
where, $\sum_{i=1}^{N}i(c_{i} -c_{i}^{\prime} )=lN$, and $l=0,1, \cdots, k$, which is from the fact that the physical states can be built by $k$ second fundamental blocks at most since $kN$ $\Psi^{1\dagger}$s always appear in the physical states. The energy eigenvalues of these states are
$E( \{ c_{i} \} ,\{ c_{i}^{\prime} \}, k)=\omega (N^{2} +kN(N-1) 
+\sum_{i=1}^{N} i(c_{i}+c_{i}^{\prime} ) )$. The $|\{ 0\} ,\{ 0 \}, k \rangle $ in (10) gives the physical ground wavefunction of our matrix model. In fact, it is the Laughlin-type wavefunction of the 4-dimensional QHF found by us.

After finishing the formal substitutions $\Psi^{1\dagger}\rightarrow 1$ and $Z^{\alpha\dagger}_{ij}\rightarrow
\delta_{ij}A_{j}^{\alpha}$ in (10), we can get $|0,k \rangle =(\epsilon^{i_{1}\cdots i_{N}}\prod_{n=1}^{N}(A^{1\dagger N-n}A^{2\dagger n-1})_{i_{n}})^{k}|0 \rangle $. Furthermore, if $u$ and $v$ are regarded as the eigenvalue
parameters of $A^{1}$ and $A^{2}$ in the coherent state picture respectively, we find that $\langle u,v |0,k\rangle =(\epsilon^{i_{1}\cdots i_{N}}\prod_{n=1}^{N}(u^{N-n}v^{n-1})_{i_{n}} )^{k} =\prod_{i<j} (u_{i}v_{j} -u_{j}v_{i})^{k}$. It is just the same as the ground state wavefunction of two-dimensional
QHF on the Haldane's spherical geometry \cite{Haldane}. However, in our system, $u$ and $v$ are the coordinates of the space ${\bf H}$ of quaternions. Conclusively, the $SU(2)$ matrix
Chern-Simons model (1) describes the 4-dimensional QHF on the ${\bf H}$. If we denote the quantum number of the $SU(2)$ representation as $J$, for the $k=1$ case, we can find that $N=2J+1$ by noticing $Z^{1\dagger}$ and $Z^{2\dagger}$ associated
with 'spin up' and 'spin down' respectively. That is, the particles of this 4-dimensional QHF are filled in the representation of the $SU(2)$ level $J$. 
For the case of general $k$, the degeneracy given by the representation of $SU(2)$ is $2J^{\prime} +1=kN$, while the particle number is still $N$. The filling factor
in this case is $\nu =N/Nk=1/k$. So this 4-dimensional QHF has the properties similar to the usual 2-dimensional QHF.

Now let us turn to the description of the matrix model corresponding to the 4-dimensional QHF of Zhang and Hu of $S^4$. In fact, this four-sphere $S^{4}$ is the base of the Hopf bundle $S^{7}$ obtained by the second Hopf fibration, from the connection\cite{Demler} between the second Hopf map and Yang's $SU(2)$ monopole\cite{Yang}. 
Equivalently, $\frac{SO(5)}{SU(2)\times SU(2)}=\frac{S^{7}}{SU(2)=S^{3}}=S^{4}$ since $S^{7}=\frac{SO(5)}{SU(2)}$. This implies that the $S^{4}$ can be identified with
the coset space of the group $SO(5)$ under the $SO(4)=SU(2)\times SU(2)$ gauge structure. Such geometrical structure plays a key role in the description of our 
matrix model. However, If one takes straightforwardly the orbital part of the $SO(5)$ generators as the generators of $SO(4)$ and decomposes them into two $SU(2)$ algebras, he can not use such $SO(4)$ blocks to generate all irreps. of $SO(5)$ \cite{Yang1}. Only if
these orbital $SO(4)$ generators are modified by the coupling of them with
a $SU(2)$ isospin , they can be used to generate the $SO(4)$ 
block states of all $SO(5)$ irreps. \cite{Yang1} 

Now it is clear that we can construct the matrix model by considering the $SO(4)$ gauge structure and some restriction conditions of the model. We express the action (1) as $S=S[Z^{\alpha}, \Psi^{\alpha}, A]$ and introduce the action of $SO(4)$ matrix Chern-Simons model as following
\begin{equation}
S_{M} =S[Z^{\alpha}, \Psi^{\alpha}, A_{0}]+S[Z^{{\bar \alpha}}, \Psi^{{\bar \alpha}}, {\bar A}_{0}] . 
\end{equation}
Two pairs of complex coordinates $Z^{\alpha}$ and $Z^{{\bar \alpha}}$ of the complex space ${\bf C}^{4}$ can be expressed as 
$Z^{1}=Q^1_1+iQ^2_1 , Z^{2}=Q^3_1+iQ^4_1$ and $Z^{{\bar 1}}=Z^{3}=Q^1_2+iQ^2_2 , Z^{{\bar 2}}=Z^{4}=Q^3_2+iQ^4_2$ in terms of eight real coordinates of the space ${\bf H}^{2}$ of quaternions equivalent to ${\bf R}^{8}$.  We can get the restricted condition of $S^{7}$ by considering the sphere
in ${\bf R}^8$, which is $ Q^{a 2}_{1} +Q^{a 2}_{2} ={\bar Z}^{a}Z^{a} =1$. $Z^{a}$, $A_{0}$ and ${\bar A}_{0}$ are finite 
$N\times N$ matrices, where $N$ is the number of particles of the system. Therefore, we write the geometrically restricted condition of the model as
\begin{equation}
Tr({\bar Z}^{a}Z^{a}) =G ,
\end{equation}
where $G$ is a parameter dependent on the model. The equations of motion of $A_{0}$ and ${\bar A}_{0}$
are taken as the constraint equations
\begin{eqnarray}
\frac{H}{2}[Z^{1},{\bar Z}^{1}]&+&\frac{H}{2}[Z^{2},{\bar Z}^{2}]+\Psi^{1}\Psi^{1\dagger}+\Psi^{2}\Psi^{2\dagger}=2H\theta, \nonumber\\
\frac{H}{2}[Z^{2},{\bar Z}^{1}]&+&\Psi^{2}\Psi^{1\dagger}=0. \nonumber\\
\frac{H}{2}[Z^{3},{\bar Z}^{3}]&+&\frac{H}{2}[Z^{4},{\bar Z}^{4}]+\Psi^{3}\Psi^{3\dagger}+\Psi^{4}\Psi^{4\dagger}=2H\theta, \nonumber\\
\frac{H}{2}[Z^{4},{\bar Z}^{3}]&+&\Psi^{4}\Psi^{3\dagger}=0.
\end{eqnarray}

From (13), one can see that the constraints
are divided into two $SU(2)$ parts of $SO(4)$. So we can use the results of the $SU(2)$ matrix Chern-Simons model. After the quantization, there exist two groups of the physical states obeying the constraint equations given by (5) and (8) with respect to $Z^{\alpha}$,$\Psi^{\alpha}$ and $Z^{\bar\alpha}$,$\Psi^{\bar \alpha}$. 
The geometrical restricted condition becomes
\begin{equation}
[Tr(Z^{a\dagger}Z^{a}) -G]|phys>_{s} =0,
\end{equation}
where, $|phys>_{s} $ represent the physical states of the geometrically stable objects. In fact, the ground wavefunctions of QHFs are such states. In general, if the physical states include the excitations, they do not satisfy the above geometrically restricted condition.

Our task is now to determine the physical states satisfying the geometrical restricted condition while obeying the Gauss constraints
and $U(1)$ charge constraints. In order to do that, we re-express the physical ground states of the $SU(2)$ model as
$[\Psi^{\alpha\dagger}Z^{\alpha\dagger}]^{j_{1} }=\epsilon^{i_{1}\cdots i_{2j_{1}+1}}
\prod_{n=0}^{2j_{1}}(\Psi^{1\dagger}Z^{1\dagger 2j_{1}-n }Z^{2\dagger n})_{i_{n+1}}$
and 
$[\Psi^{{\bar \alpha}\dagger}Z^{{\bar \alpha}\dagger}]^{j_{2} }=\epsilon^{i_{1}\cdots i_{2j_{2}+1}}
\prod_{n=0}^{2j_{2}}(\Psi^{3\dagger}Z^{3\dagger 2j_{2}-n }Z^{4\dagger n})_{i_{n+1}}$. It can be easily seen that the power of $Z^{1\dagger}$ and $Z^{2\dagger}$ in each term of $[\Psi^{\alpha\dagger}Z^{\alpha\dagger}]^{j_{1} }$ is always $2j_{1}$. If we use the direct product of the fundamental 
blocks $[\Psi^{\alpha\dagger}Z^{\alpha\dagger}]^{j_{1} }$ and $[\Psi^{{\bar \alpha}\dagger}Z^{{\bar \alpha}\dagger}]^{j_{2} }$ to construct the fundamental
block of $SO(4)$, we must make the terms $\Psi^{1\dagger}Z^{1\dagger (\cdot)}Z^{2\dagger (\cdot)}$ and
$\Psi^{3\dagger}Z^{3\dagger (\cdot)}Z^{4\dagger (\cdot)}$ become the fundamental elment
$(\Psi^{1\dagger}Z^{1\dagger m_{1}}Z^{2\dagger m_{2}}
\Psi^{3\dagger}Z^{3\dagger m_{3}}Z^{4\dagger m_{4}})_{i_{(\cdot)}}$ of the $SO(4)$ block, where $m_{1}+m_{2}=2j_{1}$ and 
$m_{3}+m_{4}=2j_{2}$. Since the fundamental blocks $[\Psi^{\alpha\dagger}Z^{\alpha\dagger}]^{j_{1} }$ and 
$[\Psi^{{\bar \alpha}\dagger}Z^{{\bar \alpha}\dagger}]^{j_{2} }$ describe the states of the full filling of degeneracy $2j_{1}+1$ and $2j_{2}+1$ of 
$SU(2)$s respectively, the direct product of them gives the degeneracy $(2j_{1}+1)\times (2j_{2}+1)$ of $SO(4)$. For the case of integer filling, each fundamental 
element of the $SO(4)$ block represents the filling of a particle, and there are $(2j_{1}+1)\times (2j_{2}+1)$ fundamental elements in this $SO(4)$ block.
Furthermore, the $(2j_{1}+2j_{2})(2j_{1}+1)\times (2j_{2}+1)-th$ power of $Z^{1\dagger}$, $Z^{2\dagger}$,$Z^{3\dagger}$ and $Z^{4\dagger}$ is included in this block. 

The physical states are built up by the factor products of all admissive $SO(4)$ blocks. Thus, the geometrical restricted condition (14)
acting on the physical states leads to
\begin{equation}
\sum (2j_{1}+2j_{2})(2j_{1}+1)\times (2j_{2}+1)=G ,
\end{equation}
where, the range of the sum can be determined by means of the fact that the $S^{7}$ is from the second Hopf fibration of $S^{4}$.
This fibration can be realized by the modified generators of orbital $SO(4)$ through the coupling of a $SU(2)$ isospin. 
If the quantum number of isospin is $I$, the smallest such irreps. of $SO(5)$ is $(2I,0)_{5}$ \cite{Yang1}, which consists of a collection of the $SO(4)$ blocks $(j_{1},j_{2})_{4}$
satisfying the relation $j_{1}+j_{2}=I, 2j_{1}=0,\cdots,2I$. This relation just gives the range of the sum (15). So we have
\begin{equation}
G=I\times \frac{1}{3}(2I+1)(2I+2)(2I+3)=2Id_{(2I,0)_{5}} .
\end{equation}
From this we get $N=d_{(2I,0)_{5}}$. For the general $k$, $G$ becomes $2kId_{(2I,0)_{5}}$.

Let us introduce the following notation
\begin{eqnarray}
[\Psi^{a\dagger}Z^{a\dagger}]^{(j_{1},j_{2})_{4}}_{i_{1}\cdots i_{d_{(j_{1},j_{2})_{4}}}}
&=& \prod_{n=0,m=0}^{2j_{1},2j_{2}}(\Psi^{1\dagger}Z^{1\dagger 2j_{1}-n}Z^{2\dagger n}\nonumber\\
&\Psi^{3\dagger}&Z^{3\dagger 2j_{2}-m}Z^{4\dagger m})_{i_{(n+1)(m+1)}} .
\end{eqnarray}
We can give straightforwardly the Fock basis of physical states of the $SO(4)$ matrix Chern-Simons model. The result is
\begin{eqnarray}
| &\{&c^a_i\},k;I \rangle 
=\prod_{j=1}^{N}(Tr Z^{1\dagger j})^{c^1_{j}}(Tr Z^{2\dagger j})^{c^2_{j}}\nonumber\\
&(&Tr Z^{3\dagger j})^{c^3_{j}}(Tr Z^{4\dagger j})^{c^4_{j}}
(\epsilon^{i_{1}\cdots i_{N=d_{(2I,0)_{5}}}} \prod_{n=0}^{I}\nonumber\\
&[&\Psi^{a\dagger}Z^{a\dagger}]^{(I-n,n)_{4}}_{i_{\sum_{m=1}^{n-1}d_{(I-m,m)_{4}}+1}\cdots i_{\sum_{m=1}^{n}d_{(I-m,m)_{4}}}})^{k}|0 \rangle ,
\end{eqnarray}
where, $\sum_{i=1}^{N}i(c_{i}^{1}-c_{i}^{2})=l_{1}(2j_{1}+1)$ for $l_{1}=0,1,\cdots, k$ and $\sum_{i=1}^{N}i(c_{i}^{3}-c_{i}^{4})=l_{2}(2j_{2}+1)$ for $l_{2}=0,1,\cdots, k$, in which $j_{1}+j_{2}=I, 2j_{1}=0,1,\cdots, 2I$. One can use the hamiltonian of this system acting on these basis to give their energy eigenvalues $E(\{c^a_i\},k;I)=\omega(2N^2+2IkN+\sum_{a,i}ic_{i}^{a})$, where $N=d_{(2I,0)_{5}}$. The physical ground state wavefunction of this matrix model is given by $| \{c^a_i =0\},k;I \rangle $ in (18). 

If we perform the formal substitution $\Psi^{a\dagger}\rightarrow 1$, $Z^{a\dagger}_{ij}\rightarrow
\delta_{ij}(\psi_{a})_{j}$ in $| \{c^a_i =0\},k;I \rangle $, and consider $\psi_{a}$ as the components of spinor reps. of $SO(5)$ in the \cite{Zhang}, we find that the ground state wavefunction of (18) in the coherent picture just becomes 
the ground state wavefunction of Zhang and Hu's 4-dimensional QHF on $S^4$\cite{Zhang}.
Hence, we have exhibited the important properties of Zhang and Hu's 4-dimensional QHF in our $SO(4)$ matrix Chern-Simons model.

In summary, we have found the new type of 4-dimensional QHF described by the matrix model (1), and established the description of the effective non-commutative field theory of Zhang and Hu's QHF provided by the action (11) together with the geometrical restricted condition (12) of $S^{7}$ which is from the second Hopf fibration of $S^{4}$.\\

{\it Note added}: An interesting work\cite{Elvang} for 4-dimensional QHE on $R^{4}$ has appeared just recently in arXiv. However, the new type of 4-dimensional QHF in the present paper is different from their QHE system.

I would like to thank Bo-Yu Hou for many valuable discussions. The work was partly supported by the NNSF of China (No.90203003), and by the Foundation of Education Ministry of China (No.010335025).


\end{document}